\documentclass{aastex}

\def\gsim{ \lower .75ex \hbox{$\sim$} \llap{\raise .27ex \hbox{$>$}} }
\def\lsim{ \lower .75ex \hbox{$\sim$} \llap{\raise .27ex \hbox{$<$}} }
\def\etal{{\it et al. }}

\title{Do clusters contain a large population of dwarf galaxies?}

\author{Carlos A. Valotto\altaffilmark{1}, Ben Moore\altaffilmark{2} 
and Diego G. Lambas\altaffilmark{1}}

\altaffiltext{1}{IATE, Observatorio Astron\'omico de C\'ordoba. 
CONICET, Argentina}
\altaffiltext{2}{Department of Physics, University of Durham, DH1 3LE, UK.}

\begin{document}

\begin{abstract}
We analyze systematic effects in the determination of the galaxy luminosity
function in clusters using a deep mock catalogue constructed from a numerical
simulation of a hierarchical universe.  The results indicate a strong tendency
to derive a rising faint end ($\alpha \lsim -1.5$) in clusters selected in two
dimensions, using a galaxy catalogue constructed with a universal flat
luminosity function with $\alpha \simeq -1.0$.  This is due to the projection
effects inherent in catalogues of clusters constructed using 2 dimensional
data.  Many of the clusters found in 2d have no significant 3d counterparts,
and most suffer from massive background contamination that cannot be corrected
for by subtracting random offset fields.  The luminosity function of high
surface brightness galaxies in the field and within small groups follows a
Schechter function with a fairly flat faint end slope, $n(L)\propto
L^{\alpha}$ with $\alpha = -0.9$ to $-1.2$.  On the contrary, observational
studies of clusters constructed using Abell, EDCC and APM catalogues are
systematically found to have steeper luminosity functions with $\alpha = -1.4$
to $-2.0$. This may be attributed to projection effects rather than a dominant
population of high surface brightness dwarf galaxies ($M\gsim M^*+2$) in
clusters. It should be straighforward to confirm our results by measuring
redshifts of these faint cluster galaxies.

\end{abstract}


\section{Introduction}

The galaxy luminosity function is one of the most direct observational probes of
galaxy formation theories. Hierarchical clustering models generically predict a
steep mass function of galactic halos that is in conflict with the field galaxy
luminosity function ({\it e.g.} Kauffmann \etal 1993, Cole \etal 1994).  The
discrepancy is resolved by invoking feedback mechanisms to suppress
star-formation and darken dwarf galaxies.  Clusters of galaxies provide the
ideal systems with which to measure the galaxy luminosity function
down to very faint magnitudes due to the large numbers of galaxies at the same
distance that can be observed within a small area on the sky (Schechter 1976).

Measuring the faint end luminosity function (hereafter LF) of galaxies in
clusters has been the subject of many detailed research papers in the last few
years ({\it e.g.} Sandage, Binggeli \& Tammann 1985; Driver \etal 1994, De
Propris \etal 1995; Gaidos 1997; Lobo \etal 1997; Lumsden \etal 1997;
Lopez-Cruz \etal 1997; Valotto et el. 1997; Wilson \etal 1997; Smith \etal
1997; Gaidos 1998, Trentham 1998a; Trentham 1998b; Driver \etal 1998; Garilli
\etal 1999).  In most of these studies, these and other authors have
demonstrated using background subtraction procedures that galaxy clusters are
dominated in number by a large population of faint galaxies.  In terms of the
faint end slope of the luminosity function, $\alpha$ lies in the range -1.4 to
-2.0.  The nature of these galaxies, their origin and evolution have important
consequences for our current understanding of galaxy formation and the
importance of environment in driving morphological evolution.

These results are at odds with observations of the field galaxy luminosity
function ({\it e.g.}  Loveday \etal 1992; Lin \etal 1997; Marzke \etal 1997;
Bromley \etal 1998, Muriel \etal 1998) which ubiquitously give a flat faint
end slope with $\alpha \approx -1.0\pm0.1$.  In other recent works, steeper LF
are found ($\alpha \simeq -1.2$) Zuca \etal 1997, Folkes \etal 1999, although
significantly flatter than the determinations in clusters.  Moreover, galaxies
within the Local Group can be observed to more than 10 magnitudes below $L_*$,
the characteristic break in the Schechter luminosity function (Schechter
1976).  Surveys of Local Group galaxies are essentially complete (Mateo 1998)
and there is no evidence for a large population of faint dwarfs - the
luminosity function is flat to $\approx L_*/10000$ (van den Bergh 1992,
Pritchet \& van den Bergh 1999).  The probability that the faint end slope in
the Local Group is as steep as 1.3 is less than 1\%.  Similarly, redshift
surveys of nearby loose groups (Ferguson \& Sandage 1990), compact groups
(Hunsberger \etal 1998) and isolated ellipticals (Mulchaey \& Zabludoff 1999)
also show evidence for flat faint end slopes (but see also Morgan \etal 1998).

The paradox is clear. In the hierarchical universe, clusters form relatively
recently from the accretion of systems similar to the Local Group.  Dynamical
processes that operate in clusters are destructive. Ram pressure stripping (Gunn
\& Gott 1972, Abadi, Moore \& Bower 1999) and gravitational tides/galaxy
harassment ({\it e.g.} Moore \etal 1996) both conspire to fade galaxies by
removing gas or stripping stars. These process are most effective for smaller,
less bound galaxies, which should manifest itself as a {\sl flattening} of the
faint end slope rather than the observed steepening.

Surface brightness selection effects complicate these issues. Apart from the
Local Group, all of the above surveys are sensitive only to high surface
brightness galaxies. Decreasing the limiting isophote for galaxy detection to
$\mu_{_{B}}\approx 27$ mag arcsec$^{-2}$ uncovers a host of previously
undetected galaxies (Impey \& Bothun 1997).  Within nearby clusters, such as
Virgo and Fornax, a divergent exponential tail to the galaxy luminosity
function cannot be ruled out (Impey etal 1988, Bothun etal 1991).  A similar
situation arises in the field surveys where we know comparatively little about
the abundance of low surface brightness objects. The observational motivation
for this paper are the observations of high surface brightness galaxies, but 
our results will be applicable to any survey where the clusters are selected 
from 2 dimensional data.

Unlike clusters which provide a large sample of galaxies at the same distance,
the field galaxy luminosity function must be determined using redshift
information.  Even with the advent of large multi-fibre instruments, the
luminosity function of cluster galaxies has not been measured out to the virial
radius, nor has it been measured to faint enough magnitudes to constrain the
faint end slope \-- both vital for comparison with cosmological models.  The
lack of large compilations of redshifts of faint galaxies imply that most
estimates of galaxy luminosity functions in clusters rely critically on
background subtraction.  The accuracy of this procedure depends on the
statistical assumption that galaxy clusters correspond to density enhancements
unbiased with respect to the distribution of foreground or background galaxies.

In spite of the lack of deep redshift surveys free from incompletness and
selection effects several groups have attempted to determine the shape of the
galaxy luminosity function from spectroscopic data in cluster fields.  By
inspection of these papers, it is apparent that flatter faint end slopes are
found over the range of magnitudes where the redshift samples are
complete. For instance, spectroscopic analysis of the central region of the
Coma cluster shows a flat faint end galaxy LF ($\alpha \simeq -1.1$) for a
single Schechter function fit, with the additional claim of a failure of a
single Schechter fit at brighter magnitudes $M \simeq -18$ to $-17.3$ (Biviano
et al. 1995).  Similarly, Koranyi et al. (1998) have analyzed a redshift
survey in the field of the poor cluster AWM 7. In this case, the galaxy LF
shows a steeper faint end ($\alpha =-1.37 \pm 0.16$), although this value is
based on a correction to the redshift data that entirely dominates the results
at faint magnitudes.

A similar analysis of A2626 and A2440 by Mohr, Geller and Wegner (1996) gives
faint end slopes $\alpha=-1.16 ^{+0.18}_{-0.16}$ and $\alpha=-1.30
^{+0.26}_{-0.20}$ (90\% confidence limits) respectively.  The study of A576 by
Mohr et al (1996) also gives a flat galaxy LF in the range of magnitudes where
the redshift survey is complete, however it steepens significantly at fainter
magnitudes where a background correction is applied.  The most complete
published redshift survey of a cluster has been published by Durret et
al. (2000) for A496. These authors show that the LF flattens at
$M_R\simeq-20.5$, where the sample is complete, and at fainter magnitudes
($-20.5< M_R <-18$), a Schechter function with $\alpha\simeq-1.2$, provides a
suitable fit to the LF with $79\%$ redshift completness.  In a study of the
galaxy LF from the Norris Survey of the Corona Borealis and A2069 supercluster
regions with spectroscopic information, Small et al. (1997) derive a flat
faint end LF ($\alpha \simeq -1.1$) consistent with the field determination.
Finally, Adami et al. (1999) analysis of spectroscopic data in the core of
Coma cluster imply that the number of low luminosity galaxies is smaller than
expected from previous estimates (Biviano et al. 1995), suggesting a flat
faint end galaxy LF.

To summarise, galaxy LF determinations using spectroscopic data suggest that
the steep faint end slope derived from background subtraction procedures
should be taken with caution. However, the spectroscopic surveys are only
complete to approximately 2 magnitudes fainter than M* whereas the photometric
surveys can probe 7 magnitudes fainter than M*.

Significant projection effects are found in Abell clusters (e.g. Lucey 1983,
Sutherland 1988, Frenk \etal 1990) that can systematically bias the observed
correlation function and the mass function of these systems.  In fact, van
Haarlem \etal (1998) find that a third of Abell clusters are not real
physically bound systems but simply projections of galaxies and groups along
the line of sight.  In this paper we use 3 dimensional mock galaxy catalogues
to construct samples of rich clusters using just 2 dimensional
information. Observing these clusters in the same way as published works
allows us to quantify how well we can recover the true luminosity function. We
will explore the biases due to foreground and background contamination, the
dependence of the results on cluster richness, and the effects of the extended
halos of clusters.

\section{Results}

\subsection{The Mock Catalogue}

Mock galaxy catalogues constructed from large N-body simulations provide ideal
data-sets with which to examine selection biases since the full 3d positional
and velocity information is available. We have used a deep galaxy catalogue
kindly constructed by Shaun Cole that is a mock realisation of the APM
catalogue.  The N-body simulation and techniques used to construct the
catalogues are discussed in detail in Cole \etal (1998), here we just review
some of the most relevant details. The N-body simulation
follows a universe dominated with a closure density of cold dark matter within
a periodic box of 345.6 $h^{-1}$ Mpc per side (where $H_0=100 h$ km s$^{-1}$
Mpc$^{-1}$ throughout).  The simulation is normalised to match the observed
number density of rich galaxy clusters such that $\sigma_8=0.55$.
``Galaxies'' are drawn from the mass distribution according to a two parameter
biasing model normalised to fit the observed r.m.s. fluctuations in the APM
catalogue in cubes of side 5 and 20 Mpc/h.

Thus, even though the initial N-body simulation is not the currently favoured
cosmological model, both the amplitude and slope of the ``galaxy'' correlation
function provide a close match to that obtained from the APM survey. Therefore, 
we do not expect our results to change if the cosmological model included a
lambda term, as currently favoured. This is of course an important point since
projection effects may depend on the degree of galaxy clustering present in
the mock catalogue.
Moreover, since the correlations between cluster and galaxy positions are
crucial in the present discussion, we have computed the cross-correlation 
function for clusters and "galaxies" in the mock catalogs as well as the cluster
autocorrelation function. The results show reasonable agreement with the observations, Lilje \& Efstathiou (1988), Croft et al. (1999), 
Dalton, Abadi, Muriel Lambas (1998)

($\xi_cg \ simeq (r/9.4)^{-1.86}$ 
($\xi_cc\ simeq (r/14.5)^{-1.96}$ ) indicating that the statistical properties 
of the galaxy distribution around the clusters in the mock catalog are 
in good agreement with the observations.
Thus, for the purpose of our analysis the mock catalog
used provides a suitable representation of
the real universe so that the results derived in this section are 
robust and not expected to change with other cosmological models that
match the observations.

The luminosities of the galaxies are drawn at random according to a Schechter
function (Schechter 1976), unbiased with respect to environment:
\begin{equation}
\phi(M)=(0.4 \ln 10) \phi^*[10^{0.4(M^*-M)}]^{1+\alpha} 
\exp[-10^{0.4(M^*-M)}].
\end{equation}
We adopt the values $\phi^*=0.014 {\rm h}^{-3} {\rm Mpc}^{-3}$, $M^*$=-19.5
and $\alpha$=-0.97 corresponding to the luminosity function of high surface
brightness field galaxies (Loveday \etal 1992).  The limiting apparent
magnitude for the mock catalogue (which subtends $\simeq 1300$ square degrees)
is $B_j=21.5$ and the final catalogues contain $1.7\times 10^6$ comparable to
those in the APM Galaxy Survey (Maddox \etal 1990a, 1990b) and
Edinburgh/Durham Southern Galaxy Catalogue (Heydon-Dumbleton, Collins, 
\& MacGillivray 1989).

\subsection{Cluster selection}

We have selected the clusters from the mock catalogue applying a similar
procedure as Lumsden \etal (1992) in the construction of the Edinburgh-Durham
cluster catalogue (EDCC). This corresponds closely to Abell's original
criteria (Abell 1958).

First, we project the galaxy catalogue on the sky and find cluster centers by
searching for galaxy overdensities.  The {\sl redshift} distance
of the cluster candidate is measured using the real velocities of the brightest
galaxies in the apparent cluster.  We then adopt a $1.5 h^{-1}$ Mpc search
radius around these centers and we consider galaxies within this radius in the
range of magnitudes $m_3$ to $m_3+2$ to define cluster richness in a similar way to
Abell and EDCC cluster identification procedures.  (Note that using this method
to determine cluster richness and distances reflects current techniques, rather
than Abell's original method that assumed the apparent magnitude of the 10th brightest
cluster galaxy was a standard candle.) The final sample of clusters for our analysis
comprises 140 objects with redshifts $z<0.15$.

We find that the typical redshift distributions in the field of the 
identified clusters in the simulations are visually comparable to the 
observed distribution of radial velocities measured for clusters. 
An example is shown in Figure 1 where we compare the redshift
distribution of a typical cluster selected in two dimensions,
panels (a) and (c), with the redshift distribution
of galaxies in the field of a cluster selected in three dimensions
with no two dimensional counterpart (see section 2.5). 
Both contain
a significant peak of galaxies at one distance that corresponds to a real
cluster, but in the 2-dim case smaller peaks are found at disparate redshifts that 
are simply groups of galaxies that lie in projection along the line of
sight to the cluster. In the absence of redshifts for all the galaxies
in the field, these groups provide a significant source of contamination
for the cluster membership.

\subsection{Luminosity function determination}

We estimate the mean galaxy luminosity function
for the total sample of clusters using only
the apparent magnitudes of galaxies and mean cluster redshifts -- the same
as published observational studies.
For a galaxy of apparent magnitude $m$ associated with a cluster 
at redshift $z$, we neglect curvature and assign an absolute magnitude 
$M=m-25-5 \log (D)$ where D is $D=z c/H_0$. 
The algorithm to determine the composite luminosity function
(Valotto \etal  1997, Lumsden \etal 1997, Muriel \etal 1998) in the absence
of galaxy radial velocities is based on a background subtraction procedure:

\begin{equation}
{\cal{N}}(M)=\frac{1}{n_{clus}}\sum_{i=1}^{n_{clus}}\frac{N_i(M)}{R_i}
\end{equation}

where $\cal{N}(M)$ is the composite luminosity function in the magnitude bin
centered in $M$ and ${N_i(M)}$ is the background-corrected number of galaxies
with magnitude $M$ of the {\it i}th cluster.  ${R_i}$ is the richness count of
the {\it i}th cluster in the range of magnitudes $-21<M<-19$ within the
projected radius $r=1.5$ h$^{-1}$Mpc and ${n_{clus}}$ is the number of
clusters contributing to the composite luminosity function.

In order to explore the galaxy LF at faint absolute magnitudes we compute the
composite LF for the nearest 22 clusters with redshift $z_{clus} < 0.07$,
using galaxies brighter than $M_{lim}=-15$.  The background de-contamination
is performed by considering a mean local background around each cluster
defined as the number of galaxies in a ring at projected radius $r_1< r_p <
r_2$. The results shown in Figure 2 correspond to $r_1=4$ Mpc and $r_2=6$ Mpc.

Figure 2 shows the resulting galaxy luminosity function in the
simulated clusters, the solid lines correspond to a best fitting Schechter
function estimated using a minimum $\chi^2$ method.  The dotted lines correspond
to the imposed Schechter function (equation 1).  The errors bars are calculated
according to Poisson statistics.  The estimated galaxy LF is significantly
steeper $\alpha = -1.41\pm 0.11$ and $M=-20.0 \pm 0.1$ than the actual LF
$\alpha=-0.97$ and $M=-19.5$. 
When we take a deeper sample of clusters surveyed
to fainter magnitudes ($M_{lim}=-16$) we find an even steeper faint end slope
$\alpha = -1.52 \pm 0.07$ and $M=-20.0 \pm 0.2$.

\subsection{Radial and richness dependence}

Can we detect any trends in how the recovered luminosity function depends on the
limiting radius used to identify clusters or the cluster richness? Smaller radii
would probe higher overdensities and would reduce the chance of background
contamination. In Figure 3 we show the results corresponding to model 1 for 
$r=0.5$ h$^{-1}$Mpc, $r=0.75$ h$^{-1}$Mpc and $r=1.0$ h$^{-1}$Mpc.
This Figure shows a similar behavior, with steeper faint end slopes than
expected, although fits to a Schechter function show a tendency for lower
values of $\alpha$ with decreasing cluster radius ({\it c.f.} Table 1).

In order to examine any possible dependence of the results with cluster richness
we have divided the cluster samples according to their richness counts, $R$,
defined in equation 2.  In Figure 4 we show the results for the samples of
different $R$ and we detect a trend of increasing negative slope as poorer
clusters are included. 

\subsection{Analysis of projection effects}

For each cluster we have considered galaxies within 1000 kms$^{-1}$ of the cluster
mean radial velocity from the mock catalogue to compute the galaxy LF.  Since in
this case we eliminate any projection effects, this tests the procedures adopted to
obtain the LF.  The results for model 1 are shown in panel (c) of Figure 5, where for
comparison we plot the imposed LF. The agreement between the input data and the
sample of real cluster galaxies is, as expected, very good (see Table 1).

We can easily examine if the systematic rise in the derived galaxy LF shown in
Figure 2 is mainly due to foreground or to background contamination. We split
the sample of cluster galaxies (identified in projection from model 1) into two: a) for each
cluster we consider only galaxies with radial velocities $V_{gal}< V_{clus}+1000
km s^{-1}$ (thus considering only foreground contamination) and b)
$V_{gal}>V_{clus}-1000 km s^{-1} $ (thus considering only background
contamination).  When we repeat the procedure to recover the LF, i.e. applying
the same de-contamination procedures described to the foreground case $V_{gal}<
V_{clus}+1000 km s^{-1}$, we find a flat galaxy LF, entirely consistent with
that originally imposed to create the galaxy catalogue.  On the other hand, a
significantly steeper LF is derived by considering galaxies with
$V_{gal}>V_{clus}-1000 km s^{-1}$ ($\alpha=-1.5$) consistent with that obtained
from the total mock catalogue (see Figure 2 and Table 1).
The results of these tests are shown in Figure 5, which demonstrate that the
artificial steep faint end slopes arises entirely due to projection of background
galaxies into the field of view of the cluster.

The observed distribution of galaxies closely matches the clustering pattern of
galaxies in our mock catalogues. We can therefore test the role of large scale
correlations and the filamentary structure of the galaxy distribution in
creating projection effects. We construct a spherically symmetric model cluster
that resembles real bound objects in the mock catalogue.  Thirteen clusters are
added to the galaxy catalogue but place in random positions with distances drawn
from the same the redshift distribution and with $z_{clus} < 0.06$. These
clusters are therefore uncorrelated with the large scale structure. We repeat
the procedure to estimate the LF using the fake clusters and we show the results
in Figure 6. Remarkably, for these clusters we recover the input luminosity
function with good precision, again demonstrating that projection effects are
dominating the bias.

We have identified clusters in 3 dimensions using the real space positions of
galaxies in the mock catalogues of the two models. We consider galaxies limited in apparent
magnitude $m=21.5$ within a radial distance of $210 h^{-1}$ Mpc
which corresponds to a
complete sample brighter than $M = -15$. To identify clusters in this
three dimensional catalogue, we use a friends-of-friends algorithm
with linking length $ l= 1.7 d_c = 0.5 h^{-1}$ Mpc. We select the most massive
systems corresponding to a mean number density of 
$2.2~ 10^{-5} {\rm h}^3 {\rm Mpc}^{-3}$,
roughly the abundance of $R>1$ Abell clusters.
We compute the galaxy LF in these samples of clusters using the same
background subtraction procedures as in section 2.3. The results
are shown in Figure 7 demonstrating that the clusters selected in
3d can be used to recover the input luminosity function.
The Schechter fit of the galaxy LF of the 3 dim cluster sample
has parameters, $\alpha=-1.08\pm0.09$ and $M^*=-19.3\pm0.2$ (table 1).

When we compare the richness of the 2d selected clusters with the 
largest physically selected 3d cluster along the same line of sight
we find a large scatter, with the richness count typically
in error by a factor of 2.

\subsection{Quantifying the effects of extended halos}

In the previous sections we have demonstrated the importance of
the correlation between cluster and galaxy positions 
in the determination of the galaxy LF parameters 
in these systems. Taking into account these results
we now consider the effects of the extended halos of clusters
on the background subtraction procedure. 
The virialised halos of clusters extend to several Mpc and the turnaround
region is much larger than this. This produces an excess of galaxies near
to the cluster that appear in projection, on the cluster core. A ``Malmquist''
like bias will cause foreground galaxies to be preferentially
``observed in the cluster''.

The estimated luminosity function $\psi(M')$ satisfies:
\begin{equation}
\psi(M')\propto\int^{\infty}_0 \phi(M) \xi_{cg}(r) R^2 dR 
\end{equation}
where $M'$ is the absolute magnitude of each galaxy assumed to be at
the mean cluster redshift.  The distances in the calculation
satisfy:
$$ r^2 = R^2 + R_o^2 - \frac{R}{R_0}(Rr_p^2/R_o)$$
where $r$ is the distance between the galaxy to the cluster center, 
$R$ is the distance of the galaxy  to the 
observer, $R_0$ is the distance of the cluster to the observer and $r_p$ is
the projected distance galaxy-cluster center.  
$\phi(M)$ is the true galaxy luminosity function for which we adopt a 
Schechter function with the same parameters as for our mock catalogue and
$\xi_{cg}(r)$ is the cluster-galaxy cross-correlation function.

We adopt a power-law model for the cluster-galaxy cross-correlation function
with parameters $r_0=10 h^{-1}$ Mpc, $\gamma=-2.0$ and an effective cutoff at
$r_{max}=40 h^{-1}$ Mpc which provides a reasonable fit to the observations
(Lilje \& Efstathiou 1988, Merchan \etal 1998).
 
The results for a typical cluster at $R_0 = 100 h^{-1}$ Mpc and $m_{lim}=21.5 $
are $\alpha=-1.0$ and $M^*=-21.8$ indicating that the extended cluster halos of
clusters do not bias the resulting LF at the faint end.  This calculation shows
that the effect of extended halos is mainly to increase the estimated value of
$M^{*}$ by the inclusion of foreground galaxies.

\section{Conclusions}

We have analyzed several sources of systematic effects present in observational
determinations of the galaxy luminosity function in clusters.  
We use a deep mock catalogue derived from a numerical simulation of a
hierarchical universe models dominated by cold dark matter to
identify clusters of galaxies in two dimensions. 
Galaxies and clusters in the mock catalog have auto-correlations and
cross-correlation functions in agreement with the observations so that
the distribution of galaxies and clusters relative to the
position of clusters resemble those actually observed. This fact 
provides confidence that our analyses on the  
mock catalog gives reliable results that apply to the real universe.  
The clusters are identified in 
projection from the galaxy distribution in a similar way as performed
to create the EDDC and
Abell catalogues. The galaxy LF in the clusters is obtained by performing 
a background subtraction procedure identical to that applied to observational
data. Our results are summarised here:

$\bullet$ \ Clusters identified in projection suffer from projection effects
that conspire to produce artificially steep faint end slopes, mimicking the
presence of a large population of dwarf galaxies.  With an input Schechter
function with a slope $\alpha=-0.97$ we recover $\alpha \approx -1.5$.

$\bullet$ \ We show that the projection effects result almost entirely from
background galaxies - many of the clusters selected in 2d have no significant
counterpart in 3d and most suffer from massive amounts of projection
that is responsible for the steep faint end slopes. 

$\bullet$ \ Unbiased estimates of mean galaxy luminosity function in clusters
may be obtained with background subtraction methods only for samples of
clusters selected in three dimensions, such as may be obtained from an X-ray
or lensing selected sample.

$\bullet$ \ We examined the effects of the extended halos and infall regions
surrounding clusters on the background subtraction procedure. Neglecting these
effects leads to a bias in the determination of $M^*$ to brighter magnitudes,
but do not affect the estimation of the faint end slope $\alpha$.

These results indicate that caution should be taken when interpreting the observed
steep faint end of the LF with a large population of faint galaxies associated with 
cluster environments \-- radial velocity data for cluster galaxies are crucial
for analyzing their faint galaxy populations and typically half of the
apparent cluster galaxies will be background objects.

We note that Abell, EDCC, and APM clusters exhibit on average a steep galaxy
LF with $\alpha \simeq -1.4$ to $-1.8$, (Valotto \etal 1997; Lumsden \etal
1997), showing that the particular choice of cluster finding algorithms in two
dimensional catalogues is not biasing the resulting LF.  On the other hand,
the observed mean galaxy LF of groups that avoid cluster neighborhoods is
found to be quite flat (Muriel, Valotto and Lambas 1998) which suggest that
either clusters have a contaminated galaxy LF or that the relative excess of
faint galaxies arises at very high densities. We also note here that the
galaxy LF is steep in irregular systems (Lopez-Cruz \etal 1997) and flat in
dense relaxed clusters, a result supported by recent spectroscopic
observations in the Coma cluster core (Adami et al. 1999).  The mounting
evidence of flat galaxy LF results from spectroscopic surveys with reasonable
completeness adds to the above considerations, and we suggest that irregular
clusters may owe their irregularity to projection effects which bias the LF to
steeper slopes.  Our results support the idea of a similar galaxy luminosity
function in clusters and the field.

\section{Acknowledgments}

The authors are especially grateful to S. Cole for providing the deep mock catalogues.
This research was supported by grants from  CONICET,
CONICOR, Agencia Nacional de Promoci\'on Cient\'{\i}fica y Tecnol\'ogica, Secretar\'{\i}a
de Ciencia y T\'ecnica
de la Universidad Nacional de C\'ordoba and Fundaci\'on Antorchas, Argentina.
BM is supported by the Royal Society.

\clearpage

\begin{deluxetable}{ccccccc}
\tabletypesize{\scriptsize}
\tablecaption{Results.}
\tablewidth{0pt}
\tablehead{
\colhead{Number of Clusters} &
\colhead{$M_{lim}$} &
\colhead{$z_{clus}$} &
\colhead{Cluster radius [Mpc h$^{-1}$]} &
\colhead{$\alpha$} &
\colhead{$M^*$} &
\colhead{Comments} 
}
\startdata

22 &-15      & 0.07 &1.50 & $-1.41\pm 0.11$ & $-20.0\pm 0.1$ & \sl{Total Sample}\\
32 &-16      & 0.08 &1.50 & $-1.52\pm 0.07$ & $-20.5\pm 0.2$ & \sl{Total Sample}\\

22 &-15      & 0.07 &1.00 & $-1.44\pm 0.12$ & $-20.1\pm 0.1$ & \\
22 &-15      & 0.07 &0.75 & $-1.31\pm 0.23$ & $-19.8\pm 0.2$ & \\

10 &-15      & 0.07 &1.50 & $-1.42\pm 0.23$ & $-20.1\pm 0.1$ &   $R\geq 40$ \\
15 &-15      & 0.07 &1.50 & $-1.29\pm 0.16$ & $-19.8\pm 0.1$ &   $R\geq 30$ \\
19 &-15      & 0.07 &1.50 & $-1.27\pm 0.12$ & $-19.8\pm 0.2$ &   $R\geq 10$ \\
 
22 &-15      & 0.07 &1.50 & $-1.10\pm 0.06$ & $-20.5\pm 0.1$ & $V_{gal}<V_{clus}+1000km/s$\\
22 &-15      & 0.07 &1.50 & $-1.61\pm 0.10$ & $-20.5\pm 0.1$ & $V_{clus}-1000km/s<V_{gal}$\\
22 &-15      & 0.07 &1.50 & $-1.07\pm 0.10$ & $-19.1\pm 0.1$ & $|V_{gal}-V_{clus}|<1000km/s$\\

13 &-14      & 0.06 &1.50 & $-0.96\pm 0.05$ & $-19.7\pm 0.1$ & \sl{Fake Clusters}\\
21 &-15      & 0.07 &1.50 & $-1.08\pm 0.09$ & $-19.3\pm 0.2$ &\sl{3-dim clusters} \\
\enddata
\end{deluxetable}

\figcaption[f1.ps]
{Redshift distribution in the fields of clusters
in the simulations.
Panels (a) and (c) correspond to a typical cluster selected in projection using 
the deep mock catalogue. For comparison,
panels (b) and (d) show the observed distribution of galaxy redshifts 
in the field of a cluster selected in three dimensions with no
2-dim counterpart in the same catalog
(see section 2.5).}

\figcaption[f2.ps]
{ Composite cluster luminosity function derived from the mock catalog.
(See section 2.3 for details on the normalisation of the
cluster galaxy luminosity function.)
The best fitting Schechter function is shown as a solid line ($\alpha=-1.52$ and
$M_*=-20.5$).  The dashed line shows the original input luminosity function used
to construct the mock catalogue ($\alpha=-0.97$ and $M_*=-19.5$).  }

\figcaption[f3.ps]
{The composite cluster luminosity function for different cluster radii:
a) r=1.0 Mpc h$^{-1}$, b) r=0.75 Mpc h$^{-1}$, c) r=0.5 Mpc h$^{-1}$,
d) r=0.25 Mpc h$^{-1}$. 
Solid lines correspond to the fit to the projected data shown in
Figure 2.}

\figcaption[f4.ps]
{The luminosity function of cluster galaxies of different richness:
a) $R\geq 40$, b) $R\geq 30$ and c) $R\geq 10$. 
The solid curves correspond to the  fit shown in Figure 2.} 

\figcaption[f5.ps]
{Composite luminosity function for three different cuts in the 
redshift distribution. 
a) $V_{gal}\leq V_{clus}+1000 km/s$
b) $V_{gal}\geq V_{clus}-1000 km/s$.
c) $V_{clus}-1000 km/s\leq V_{gal}\leq V_{clus}+1000 km/s$,
The solid curves correspond to the fit shown in Figure 2 and the dashed
curve shows the input galaxy luminosity function.} 

\figcaption[f6.ps]
{Composite luminosity function for clusters with positions 
uncorrelated with large scale structure. 
The solid line shows the true galaxy LF ($\alpha=-0.97$, $M^*=-19.5$)}

\figcaption[f7.ps]
{Composite galaxy LF of real clusters identified in 3
dimensions (filled circles). For comparison we show the composite galaxy 
luminosity function of clusters identified in 2 dimensions 
shown in Figure 2 (open circles).
} 

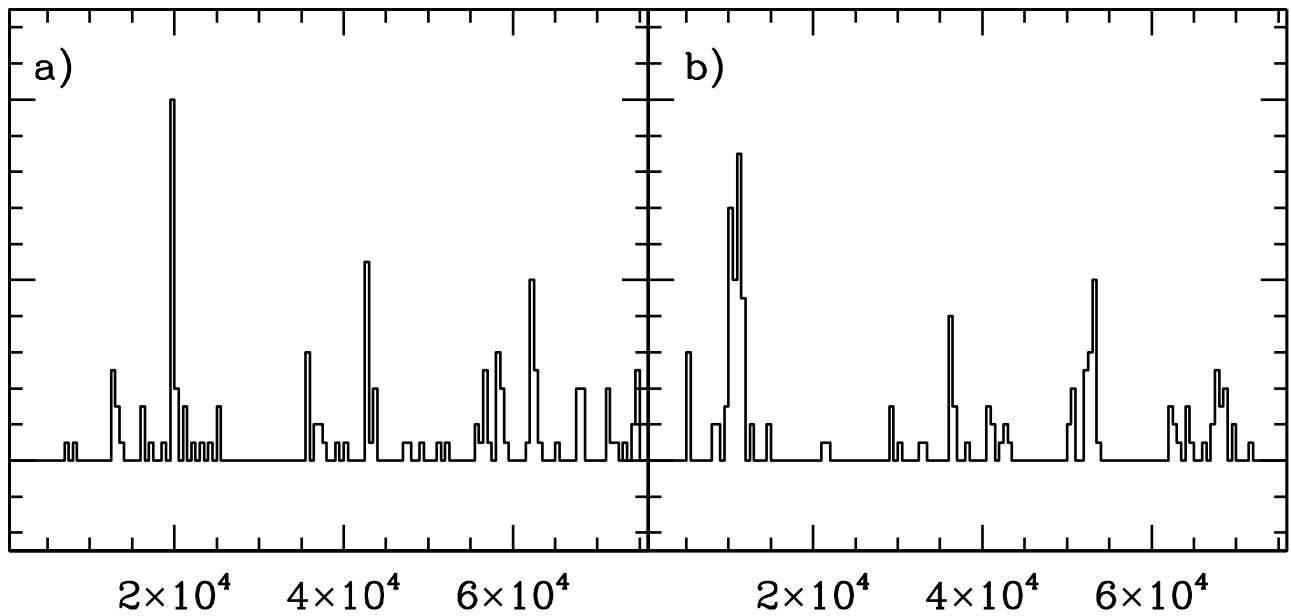
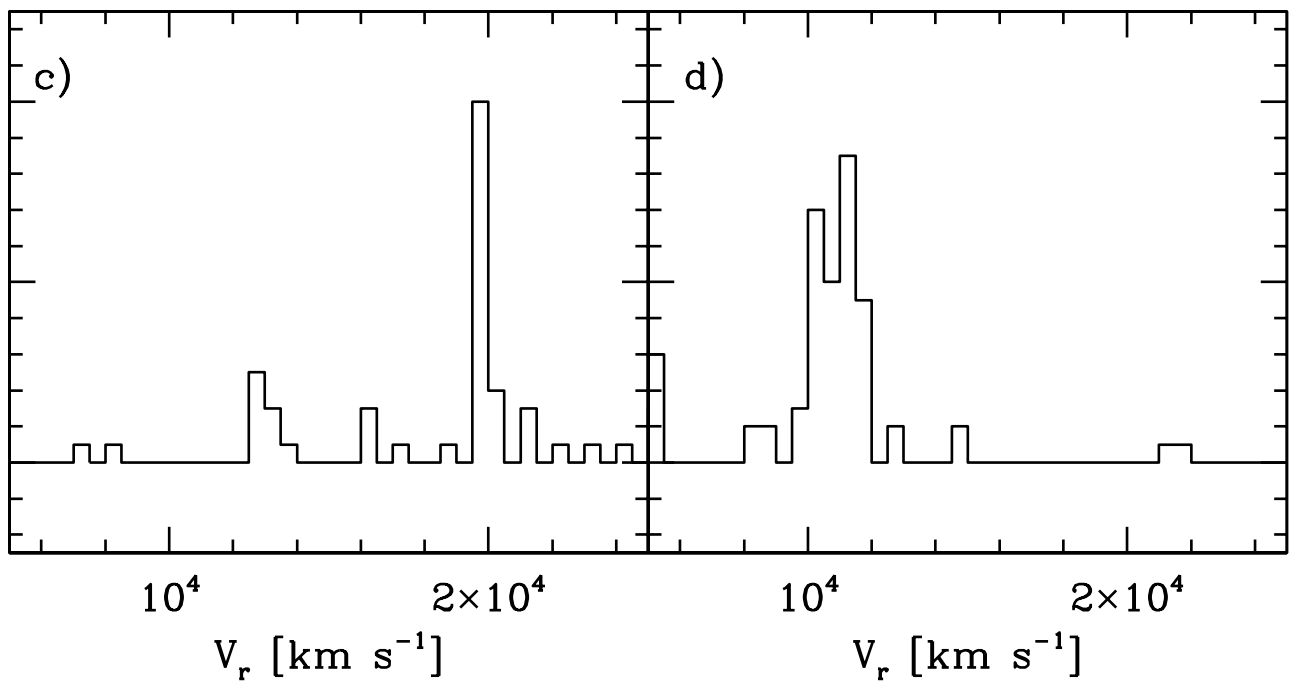
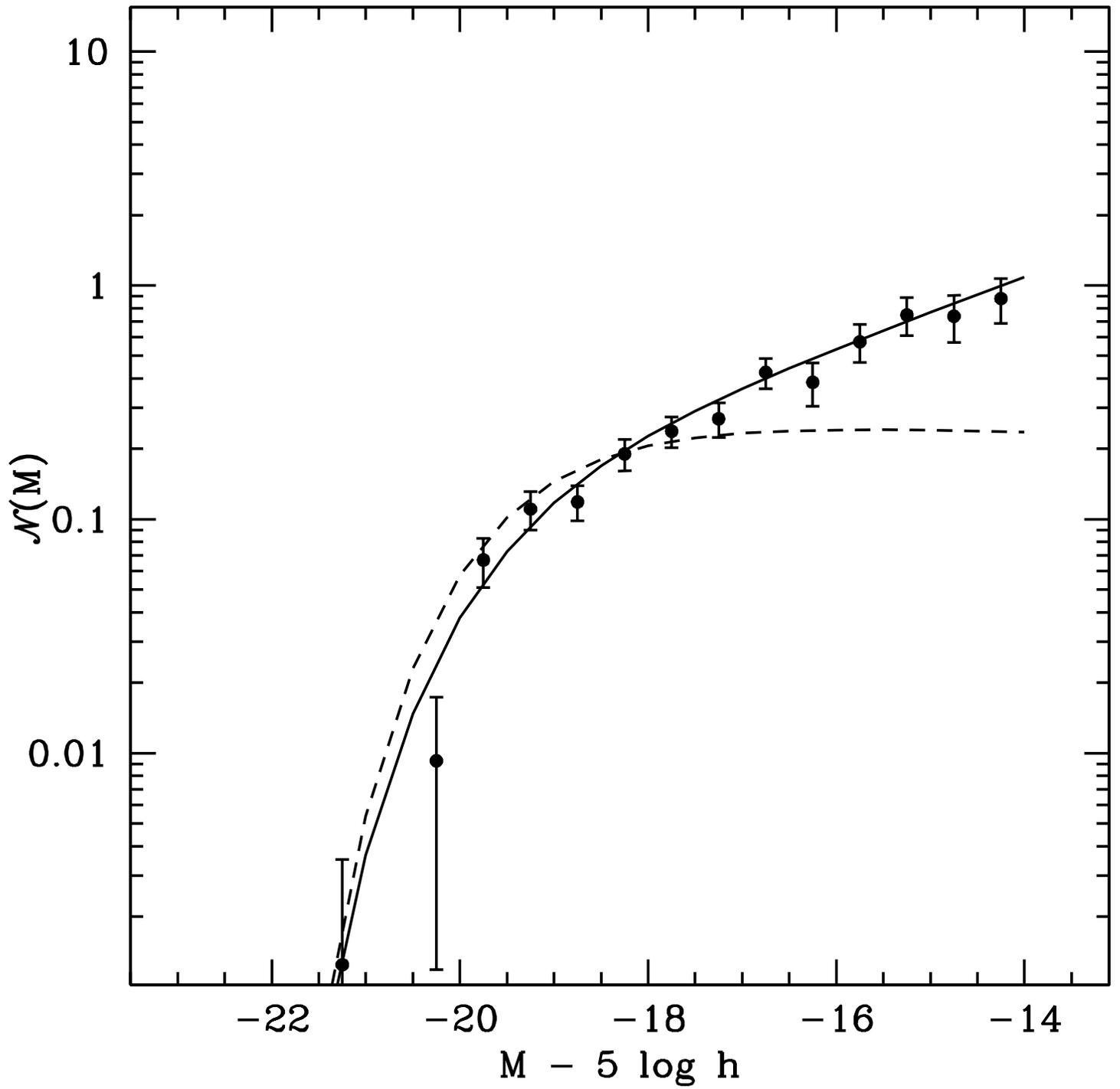
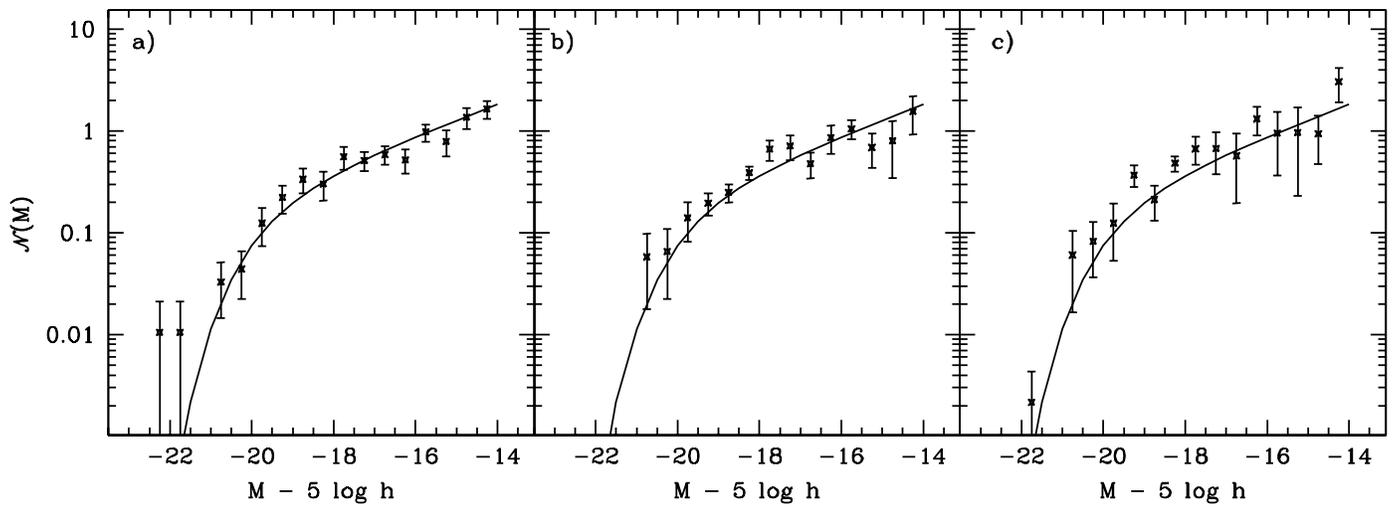
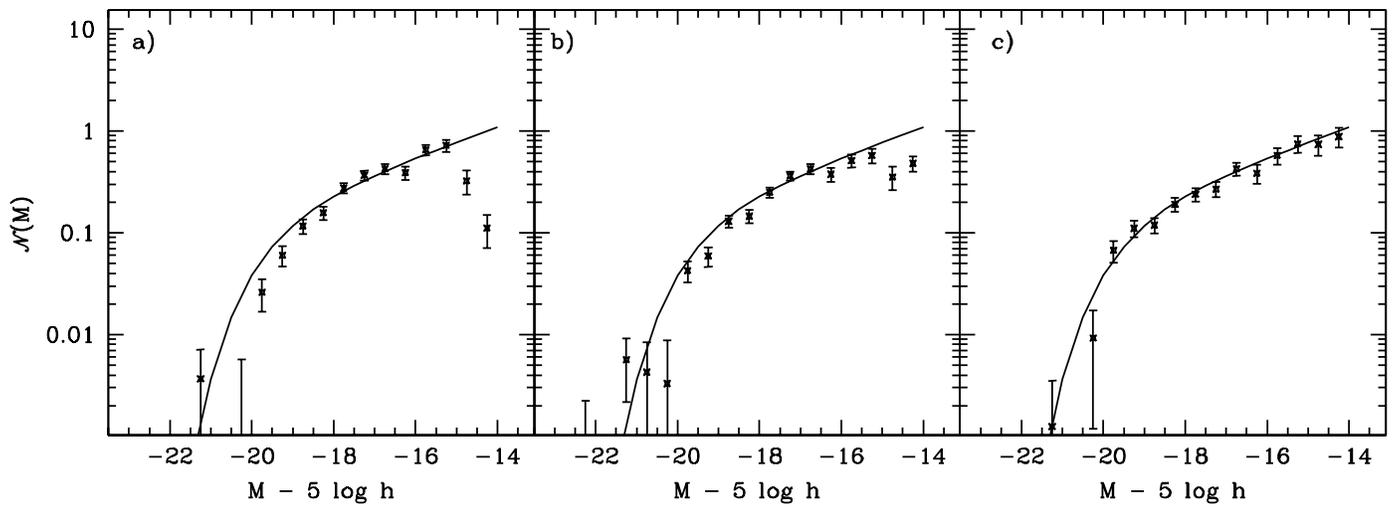
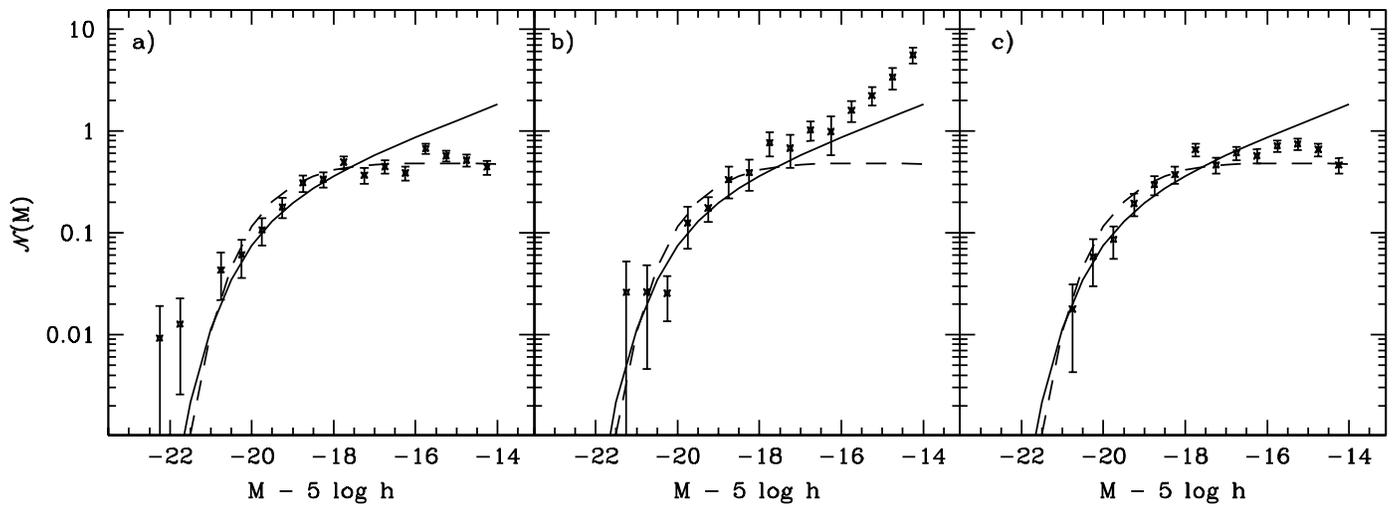
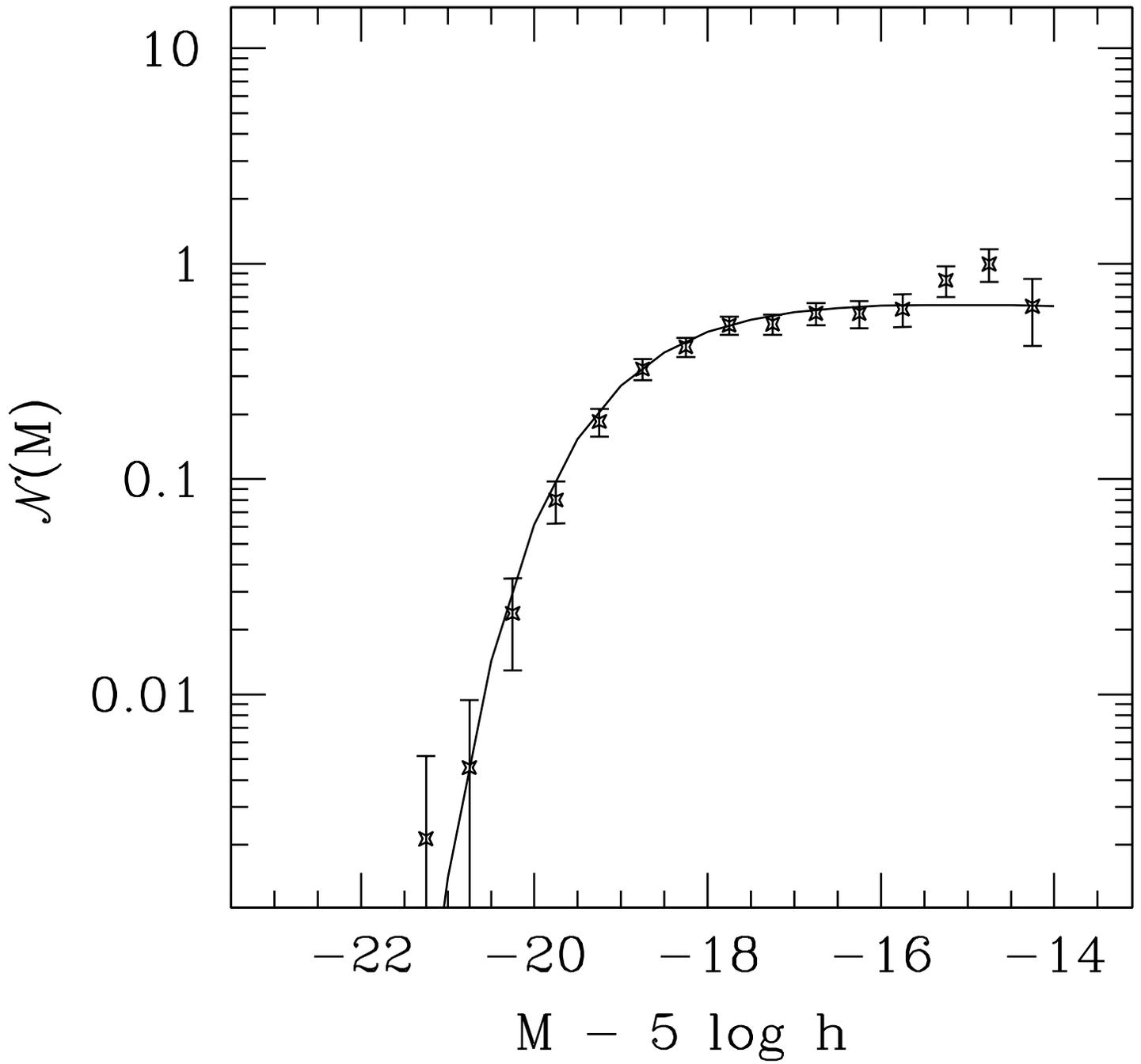
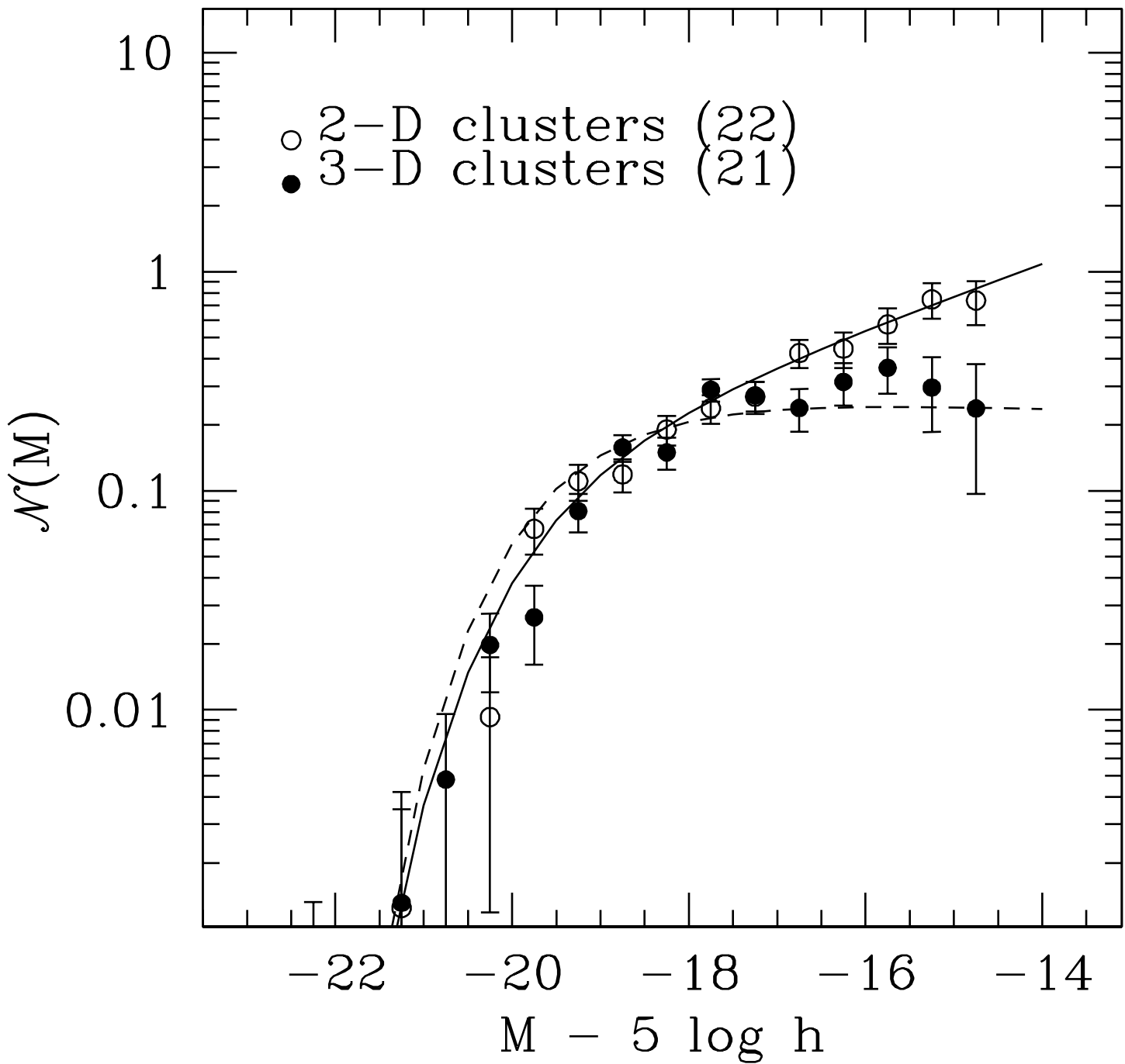
\end{document}